\documentclass[twocolumn,a4page]{sn-jnl}

\usepackage{amsmath}
\usepackage{amssymb}
\usepackage{mathrsfs}

\usepackage{xcolor}
\usepackage{graphicx}
\usepackage{dcolumn}
\usepackage{bm}
\usepackage{booktabs}
\usepackage{multirow}
\usepackage{xspace}
\usepackage{paralist} 
\usepackage{amsmath}
\usepackage{nicefrac}
\usepackage{hyperref}
\usepackage{cite}

\newcommand{\Qsq}{\ensuremath{Q^{2}}\xspace}
\newcommand{\xbj}{\ensuremath{x_\text{Bj}}\xspace}

\newcommand{\pt}{\ensuremath{p_\text{T}}\xspace}

\newcommand{\GeVsq}{\ensuremath{\mathrm{GeV}^2} }
\newcommand{\GeV}{\ensuremath{\mathrm{GeV}} }

\newcommand{\as}{\ensuremath{\alpha_{\rm s}}\xspace}

\newlength{\dinwidth}
\newlength{\dinmargin}
\begin{document}

\received{March 2024}
\printhistory
\motto{}{Published in EPJC Letters \hfill DESY-24-034}

\title{Observation and differential cross section measurement of neutral current DIS events with an empty hemisphere in the Breit frame} 

\subtitle{H1 Collaboration}

\author[50]{\fnm{V.}\sur{Andreev}}
\author[34]{\fnm{M.}\sur{Arratia}}
\author[46]{\fnm{A.}\sur{Baghdasaryan}}
\author[9]{\fnm{A.}\sur{Baty}}
\author[40]{\fnm{K.}\sur{Begzsuren}}
\author[17]{\fnm{A.}\sur{Bolz}}
\author[30]{\fnm{V.}\sur{Boudry}}
\author[15]{\fnm{G.}\sur{Brandt}}
\author[26]{\fnm{D.}\sur{Britzger}}
\author[7]{\fnm{A.}\sur{Buniatyan}}
% ...
\affil[7]{\orgaddress{School of Physics and Astronomy, University of Birmingham, Birmingham, United Kingdom$^{f3}$}}
% ...
\affil[9]{\orgaddress{University of Illinois, Chicago, IL 60607, USA}}
% ...
\affil[15]{\orgaddress{II. Physikalisches Institut, Universität Göttingen, Göttingen, Germany}}
% ...
\affil[17]{\orgaddress{Deutsches Elektronen-Synchrotron DESY, Hamburg and Zeuthen, Germany}}
% ...
\affil[26]{\orgaddress{Max-Planck-Institut für Physik, München, Germany}}
% ...
\affil[30]{\orgaddress{LLR, Ecole Polytechnique, CNRS/IN2P3, Palaiseau, France}}
% ...
\affil[34]{\orgaddress{University of California, Riverside, CA 92521, USA}}
% ...
\affil[40]{\orgaddress{Institute of Physics and Technology of the Mongolian Academy of Sciences, Ulaanbaatar, Mongolia}}
% ...
\affil[46]{\orgaddress{Yerevan Physics Institute, Yerevan, Armenia}}
% ...
\affil[50]{\orgaddress{Affiliated with an institute covered by a current or former collaboration agreement with DESY}}

\abstract{\unboldmath%
  The Breit frame provides a natural frame to analyze lepton--proton scattering events. In this reference frame, the parton model hard interactions between a quark and an exchanged boson defines the coordinate system such that the struck quark is back-scattered along the virtual photon momentum direction. In Quantum Chromodynamics (QCD), higher order perturbative or non-perturbative effects can change this picture drastically. 
As Bjorken-$x$ decreases below one half, 
  a rather peculiar event signature is predicted with increasing probability, where no radiation is present 
  in one of the two Breit-frame hemispheres and all emissions are to be found in the other hemisphere.
  At higher orders in $\alpha_s$ or in the presence of soft QCD effects, predictions of the rate of these events are far from trivial, and that motivates measurements with real data.
  We report on the first observation of the empty current hemisphere events in electron--proton collisions at the HERA collider using data recorded with the H1 detector at a center-of-mass energy of 319\,GeV.
  The fraction of inclusive neutral-current DIS events with an empty hemisphere is found to be
  $0.0112 \pm 3.9\,\%_\text{stat} \pm 4.5\,\%_\text{syst} \pm 1.6\,\%_\text{mod}$ in the selected kinematic region of $150<\Qsq<1500\,\GeVsq$ and inelasticity $0.14<y<0.7$.
  The data sample corresponds to an integrated luminosity of 351.1\,pb$^{-1}$, sufficient to enable differential cross section measurements of these events.
  The results show an enhanced discriminating power at lower Bjorken-$x$ among different Monte Carlo event generator predictions. 
}

\maketitle

\section{Introduction \label{sec:intro}}
\begin{figure}[htb!]
  \centering
  \includegraphics[width=0.90\linewidth,trim={ 0 0 0 0},clip]{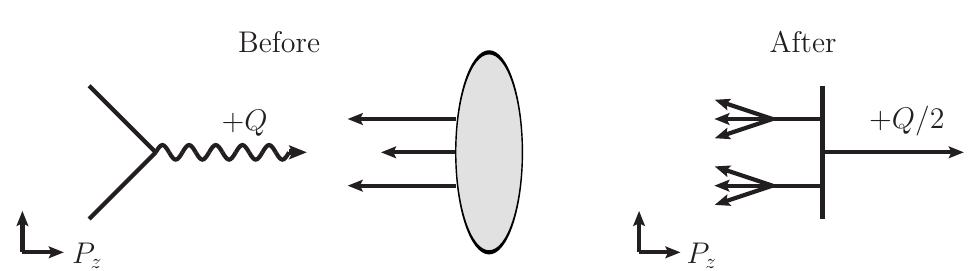}
\vspace{0.4cm}
  \includegraphics[width=0.95\linewidth,trim={ 0 0 0 0},clip]{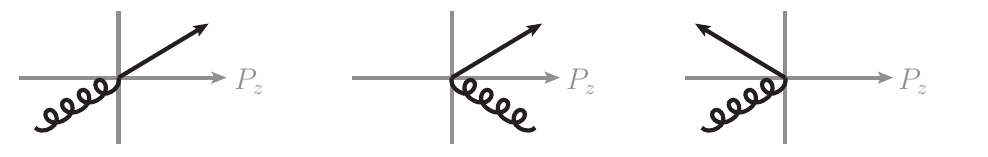}
  \caption{
    Illustration of the Breit frame. Top: Parton configuration before and after the absorption of the virtual photon.
    Bottom: Possible quark and gluon configurations at $\mathcal{O}(\as)$ after the interaction with the
    virtual photon in the Breit frame.
  }
  \label{fig:Breit}
\end{figure}

Lepton--nucleus scattering played an important role in establishing
Quantum
Chromodynamics~\cite{Fritzsch:1973pi,Gross:1973ju,Gross:1973id,Politzer:1973fx,TaylorKendallFriedman:1991,Abramowicz:1998ii,Klein:2008di,Newman:2013ada,Gross:2022hyw}
as the theory of strong interactions, and it receives continued
attention~\cite{Begel:2022kwp,vanBeekveld:2023lfu,Knobbe:2023ehi,Lappi:2023otn,Banfi:2023mhz,Cao:2024ota} due
to its sensitivity to interesting emergent QCD phenomena.  
Neutral-current deep-inelastic scattering (NC DIS) 
is mediated by a virtual electroweak boson. At low energies the
process can be considered as a photon--hadron interaction~\cite{Feynman:1973xc}.
While experiments record data in the laboratory rest frame,
a natural way to study photon--hadron interactions is given by the
Breit frame~\cite{Feynman:1973xc,Streng:1979pv}, where
the virtual
photon moves along the $z$ axis with momentum $Q$ and the proton
fragments into the opposite hemisphere. 
The space-like photon four-momentum in that frame reads $q_b=(0;0,0,Q)$ with
its Lorentz-invariant virtuality $Q^2=-q_b\cdot q_b$.
In the naive parton model only the process $\gamma^\ast + q\to q$
contributes, where $\gamma^\ast$ is the interacting virtual photon and $q$ is the (struck) quark. The struck quark moves with momentum $+\frac{Q}{2}$ along the positive $z$ direction in the Breit frame, i.e.\ opposite to the remnants of the broken-up proton. 
An illustration of the Breit frame is displayed in Fig.~\ref{fig:Breit}.

In the Breit frame, the sign of the $z$ component of momentum defines two hemispheres:
the positive half is referred to as \emph{current
hemisphere}, and the negative as \emph{target hemisphere}, also known as \emph{fragmentation hemisphere}.
In leading-order perturbative QCD (pQCD), the current hemisphere is 
analogous to a single hemisphere of an $e^+e^-\to q\bar{q}$ final state at $\sqrt{s_{e^+e^-}}=Q$,
where $\sqrt{s_{e^+e^-}}$ is the $e^+e^-$ center-of-mass energy~\cite{Webber:1995ka}. 

While in the quark-parton model picture the struck quark is simply backscattered into the current hemisphere with longitudinal momentum $\tfrac{Q}{2}$, new event topologies emerge when QCD effects are considered. In pQCD, when two or more final-state partons are present, the Breit frame and the partonic center-of-mass frame deviate from each other.
At order $\alpha_s$, there are two hard processes to consider:
Gluon bremsstrahlung $\gamma^\ast + q \to q + g$, and quark-pair production $\gamma^\ast + g \to q + \bar{q}$, where the interacting gluon in the initial state is a constituent parton of the incoming hadron.
As shown in Fig.~\ref{fig:Breit}, the two outgoing partons in these processes balance in the transverse plane, while the momentum sum in the longitudinal direction can be different from $\tfrac{Q}{2}$.
%
For $\xbj<0.5$ configurations are kinematically accessible where all outgoing partons are on-shell and are scattered into the target hemisphere, such that the \emph{current hemisphere remains empty}~\cite{Streng:1979pv,Dasgupta:1997ex,Kang:2014qba}.
Qualitatively, this topology can be explained by an off-shell parton with energy fraction $x>\xbj$ entering the hard interaction, then 
producing a massive dijet system.
The larger $x$ than \xbj, where the latter defines the Breit frame, the more is the the dijet system boosted into the beam hemisphere of the Breit frame, possibly leaving the current hemisphere empty. 
The probability of this event configuration increases as \xbj is decreasing below a value of $\frac{1}{2}$.

When considering even higher order corrections in pQCD with massless partons, the number of empty hemisphere events will successively decrease with
the order in $\alpha_s$, since the probability of (infinitly soft) radiation into the current hemisphere rises.
On the other hand, QCD confinement demands that the massless partons fragment into a finite number of massive particles, a process which is difficult to predict in great detail. 
Consequently, an accurate prediction of empty hemisphere events in higher order pQCD or at the  
particle level is considerably less trivial than in the two parton picture.\\
\indent In this letter we present the first measurement of the production rate and differential distributions of events with an
empty current hemisphere in the Breit frame using
data recorded in electron--proton and positron--proton collisions.

\section{Experimental method}
Data were recorded with the H1 detector~\cite{H1:1993ids,H1CalorimeterGroup:1993boq,H1:1996prr,H1:1996jzy,H1SPACALGroup:1996ziw,Pitzl:2000wz} at the HERA collider at DESY
in the years 2003 to 2007, where electron or positron beams with
energy of 27.6\,GeV 
collided with a proton beam of $E_p=920\,\GeV$, resulting in a center-of-mass
energy of $\sqrt{s}=319\,\GeV$.
The analyzed data sample corresponds to an integrated luminosity of
351.1\,pb$^{-1}$~\cite{H1:2012wor}, of which about one half was
taken with an electron beam, and the other half with a positron beam.
The H1 experiment consists of a series of subdetectors positioned
around the $ep$ interaction point, including 
tracking detectors, calorimeters and the muon system.
The central tracking system covers the polar angular range
$15^\circ<\theta<165^\circ$ and consists of silicon vertex detectors,
drift and proportional chambers.
Charged particles are measured with a transverse momentum resolution
of $\sigma_{p_\text{T}}/p_\text{T}=0.2\%p_\text{T}/\GeV\oplus1.5\%$.
Two main calorimeters are used: a liquid argon calorimeter (LAr)
and a lead-scintillating fiber calorimeter (Spacal).
The LAr consists of electromagnetic sections made of lead absorbers and hadronic sections with steel absorbers.
It covers the range $4^\circ<\theta<154^\circ$ and full azimuthal angle.
Its energy resolution
is $\sigma_E /E = 11\%/\sqrt{E/\GeV} \oplus 1\%$
for electrons and $\sigma_E /E \simeq 55\%/ \sqrt{E/\GeV} \oplus 3\%$ for
charged pions.
The Spacal consists of an electromagnetic and hadronic section and
records the energy deposited in the backward direction in the range $153^\circ<\theta<177.5^\circ$.
In this analysis, 
it is used to measure the hadronic energy flow in the region.
The tracking detectors and calorimeters are surrounded by a
superconducting solenoid providing a magnetic field of 1.16\,T.
The return yoke is equipped with a muon system.

The analysis strategy follows closely event selection and
methodologies used previously~\cite{H1:2012qti,H1:2014cbm}.
The trigger requires an energy deposit in the electromagnetic section of the LAr.
This trigger requirement limits the measurements to the kinematic region of $\Qsq\gtrsim150\,\GeVsq$ and inelasticity $y<0.7$.
The scattered lepton candidate is identified by requiring that the electromagnetic cluster with the highest momentum is matched to a track and its energy exceeds
$E_{e^\prime}>11\,\GeV$.
A set of fiducial cuts and quality requirements~\cite{H1:2012qti,H1:2014cbm} suppress non-collision backgrounds and backgrounds from
processes other than neutral-current DIS such as photoproduction,
charged-current DIS, or QED Compton.
The electromagnetic energy scale is calibrated in situ
using the double-angle reconstruction method to predict the electron energy.
The hadronic energy scale is calibrated by exploiting the
transverse momentum balance of the lepton and hadronic final state using
a dedicated selection of NC DIS events~\cite{Kogler:2011zz}.
An isolated electromagnetic energy deposit without a track pointing to
it is classified as a radiated photon from the lepton vertex. It is combined with the lepton candidate if the angular distance is closer to the lepton than to the beam axis, or it is removed from the list of reconstructed particle candidates.
After removing tracks and clusters that are associated with the
scattered lepton candidate~\cite{H1:2003xoe}, a particle-flow algorithm~\cite{energyflowthesis,energyflowthesis2,energyflowthesis3} is applied to
combine information from the calorimeter and tracking systems and
define the objects of the hadronic final state (HFS).
The total longitudinal energy-momentum balance of all recorded
particle candidates ($\Sigma_\text{tot}=\sum_i(E_i-P_{z,i})$) is required to be in the
range $45<\Sigma_\text{tot}<62\,\GeV$, 
which suppresses events with hard initial state QED radiation and
contributions from photoproduction. This ensures an optimal resolution is achieved of the DIS kinematic observables for the selected events.

The kinematic quantities of the DIS process, i.e.~the inelasticity $y$, the virtuality of the
exchanged boson \Qsq, and the Bjorken variable
\xbj
are reconstructed using the I$\Sigma$ reconstruction method~\cite{Bassler:1994uq} as
\begin{equation}
  y = \frac{\Sigma_h}{\Sigma_\text{tot}}~,~~~~~
  \Qsq = \frac{P_{\text{T},e^\prime}^2}{1-y} ~,~~~~~
  \xbj E_p = \frac{Q^2}{2\Sigma_h} \,,\nonumber
\end{equation}
where $\Sigma_h$ is the longitudinal energy momentum balance of the
HFS, expressed as $\sum_{i\in\text{HFS}}(E_i-P_{z,i})$.
The event selection is restricted to the region $\Qsq>150\,\GeVsq$ and
$0.14<y<0.7$.

The Lorentz transformation to the Breit frame is
given by the boost vector
$\vec{b}=(\frac{b_x}{b_E},\frac{b_y}{b_E},\frac{b_z}{b_E})$ using the 
Lorentz four-vector
\begin{equation}
  \label{eq:b}
  b=
  q-\frac{q\cdot q}{q\cdot\hat{z}}\hat{z} =
  q+2\xbj P\,,
\end{equation}
where $\hat{z}=(1;0,0,1)$, $q$ is the photon four-momentum, $P$ is the proton (beam) four-momentum in the laboratory rest frame and the proton mass is negligible. 
A convenient  expression for the boost to the Breit frame in matrix notation is presented in Appendix~\ref{sec:boost}.

\begin{figure}[htb!]
  \centering
  \includegraphics[width=0.60\linewidth,trim={300 110 30 110 },clip]{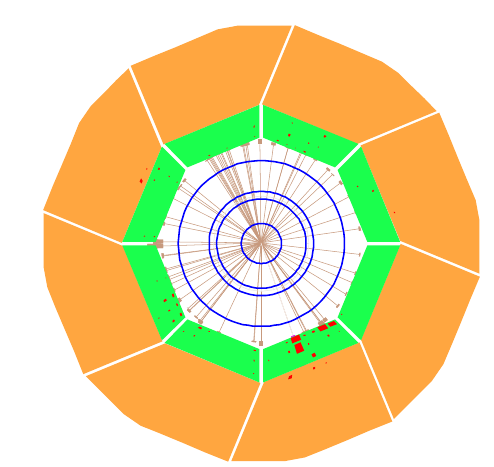}
\includegraphics[width=0.95\linewidth,trim={ 0  130 290 130 },clip]{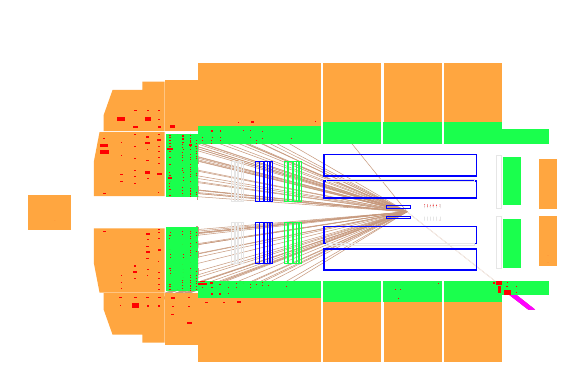}
  \caption{
    Display of an event with $\Qsq=295\,\GeVsq$, $y=0.18$, and
    $\xbj=0.015$ in the H1 detector in the radial view and in the longitudinal side view.
    The colored areas indicate different subdetector components, the red
    boxes indicate the energies of calorimeter clusters, and the straight lines show reconstructed particle candidates. } 
  \label{fig:event}
\end{figure}

Events with an empty current hemisphere are identified by requiring
that no HFS particle candidate has positive longitudinal momentum in the
Breit frame, i.e.\ 
\begin{equation}
  q\cdot p_i < 0 ~~~~ \forall ~ i\in\text{HFS}\,,
\end{equation}
and consequently the observed total energy in the current hemisphere is zero,
\begin{equation}
  E_\mathcal{C}=0\,.
\end{equation}
In order to obtain a higher purity of the empty hemisphere event
sample, the photon four-vector $q$ is reconstructed
from the observed final state as $q=k-k^\prime$, where $k^\prime$ is the scattered lepton momentum and $k$ the lepton beam momentum four-vector. The effective lepton beam momentum is reconstructed as 
$k=(\tfrac{\Sigma_\text{tot}}{2};0,0, \tfrac{\Sigma_\text{tot}}{2})$, which reduces sensitivity to initial state QED radiation as compared to using the nominal beam momentum. 
The selected empty hemisphere events
often have characteristic signatures with high particle multiplicities
in the forward region. An example event is illustrated
in Fig.~\ref{fig:event}. 

Monte Carlo (MC) event simulations are employed to correct the data for
detector acceptance and resolution effects.
The signal NC DIS events are generated using the two standard programs
Djangoh~1.4~\cite{Charchula:1994kf} and
Rapgap~3.1~\cite{Jung:1993gf} as used in H1.
Both generators use LO matrix elements which include
diagrams for boson-gluon fusion and QCD-Compton processes. Higher
order processes are included in Djangoh via the implementation of the
Color Dipole Model in Ariadne~\cite{Lonnblad:1992tz}. In Rapgap they are included via parton showers in the leading logarithmic approximation.
The non-perturbative components are modelled with the Lund
hadronization model as implemented in Pythia~6~\cite{Andersson:1983ia,Sjostrand:1993yb,Sjostrand:1994kzr,Sjostrand:2001yu} with parameters optimized by the ALEPH
Collaboration~\cite{ALEPH:1996oqp}.
Both models use the CTEQ6L PDF set~\cite{Pumplin:2002vw}. 
Additionally, both generators adopt Heracles~\cite{Kwiatkowski:1990es} for
the simulation of higher order QED radiative effects.
The particle level is set by particles with proper lifetime
$c\tau > 10\,$mm, whereas particles with lower lifetime are included by 
their decay products.
The detector effects on the simulated particle-level events are processed with a detailed simulation of the
H1 detector, which is based on Geant~3~\cite{Brun:1987ma} and fast
shower simulation programs~\cite{Fesefeldt:1985yw,Grindhammer:1989zg,Gayler:1991cr,Kuhlen:1992ey,Grindhammer:1993kw,Glazov:2010zza}.
The simulated data are processed with the same offline analysis
chain as the real data~\cite{Britzger:2021xcx}.

\begin{figure}[htb!]
   \includegraphics[width=0.99\linewidth,trim={10 0 30 0 },clip]{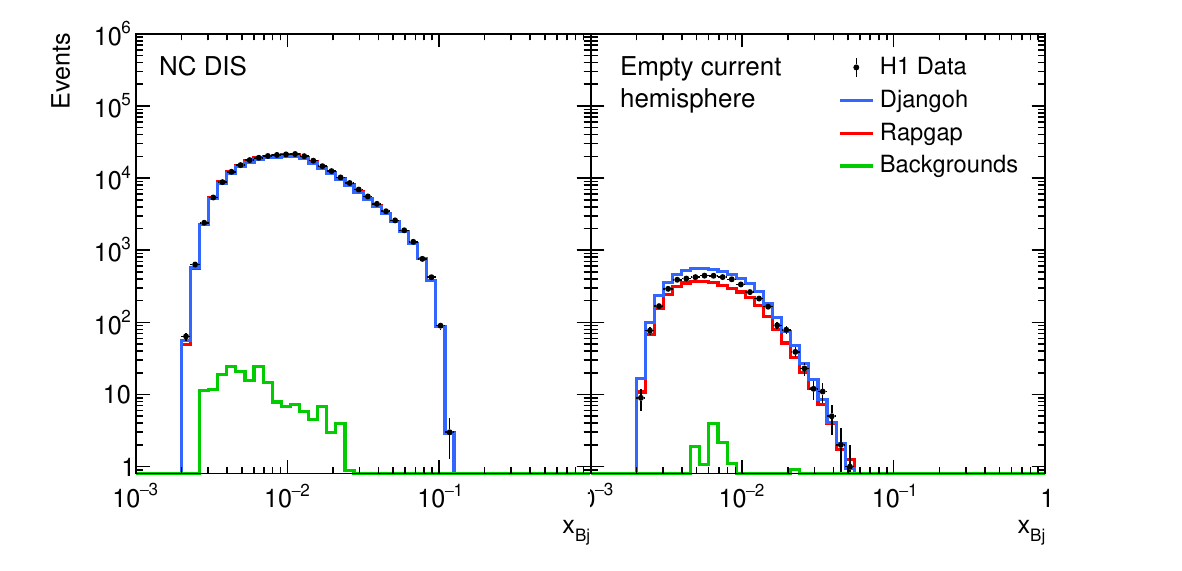}
 \includegraphics[width=0.99\linewidth,trim={10 0 30 0 },clip]{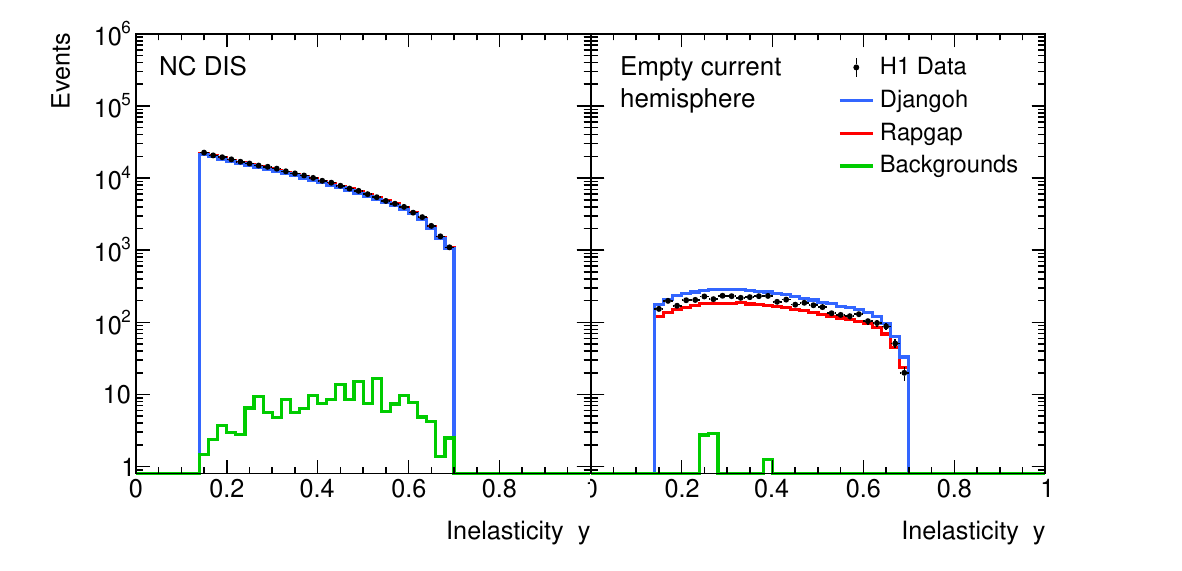}  \\
   \includegraphics[width=0.99\linewidth,trim={10 0 30 0 },clip]{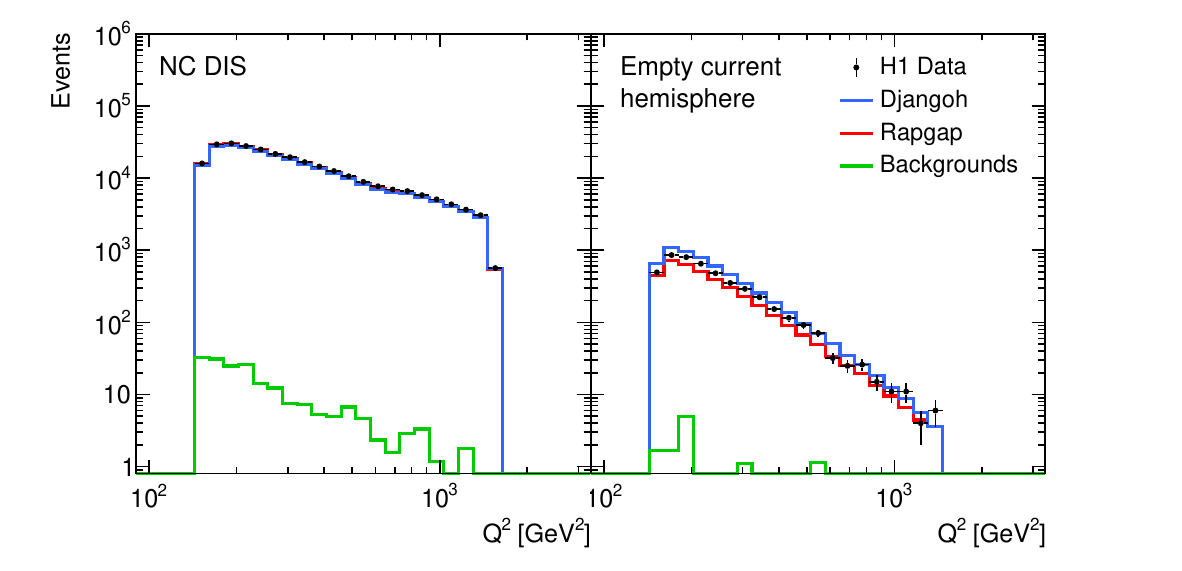}\\
  \caption{
    Event distributions as functions of \xbj\,(top), $y$\,(middle), and \Qsq\,(bottom) at
    the detector level in comparison with the simulated MC event
    samples from Djangoh and Rapgap that further include background events from
    photoproduction and lower \Qsq.
    The left panels display events after selection of all NC DIS
    events, and the right panels display events with an empty current
    hemisphere in the Breit frame.
  }
  \label{fig:controlplots}
\end{figure}

The reconstructed kinematic observables $\xbj$, $y$ and \Qsq\ are
compared with MC predictions in Fig.~\ref{fig:controlplots} for
all NC events in the selected phase space and for events with a
reconstructed empty current hemisphere.
Good agreement between the simulations and data are observed.

\begin{figure}[htb!]
  \includegraphics[width=0.62\linewidth]{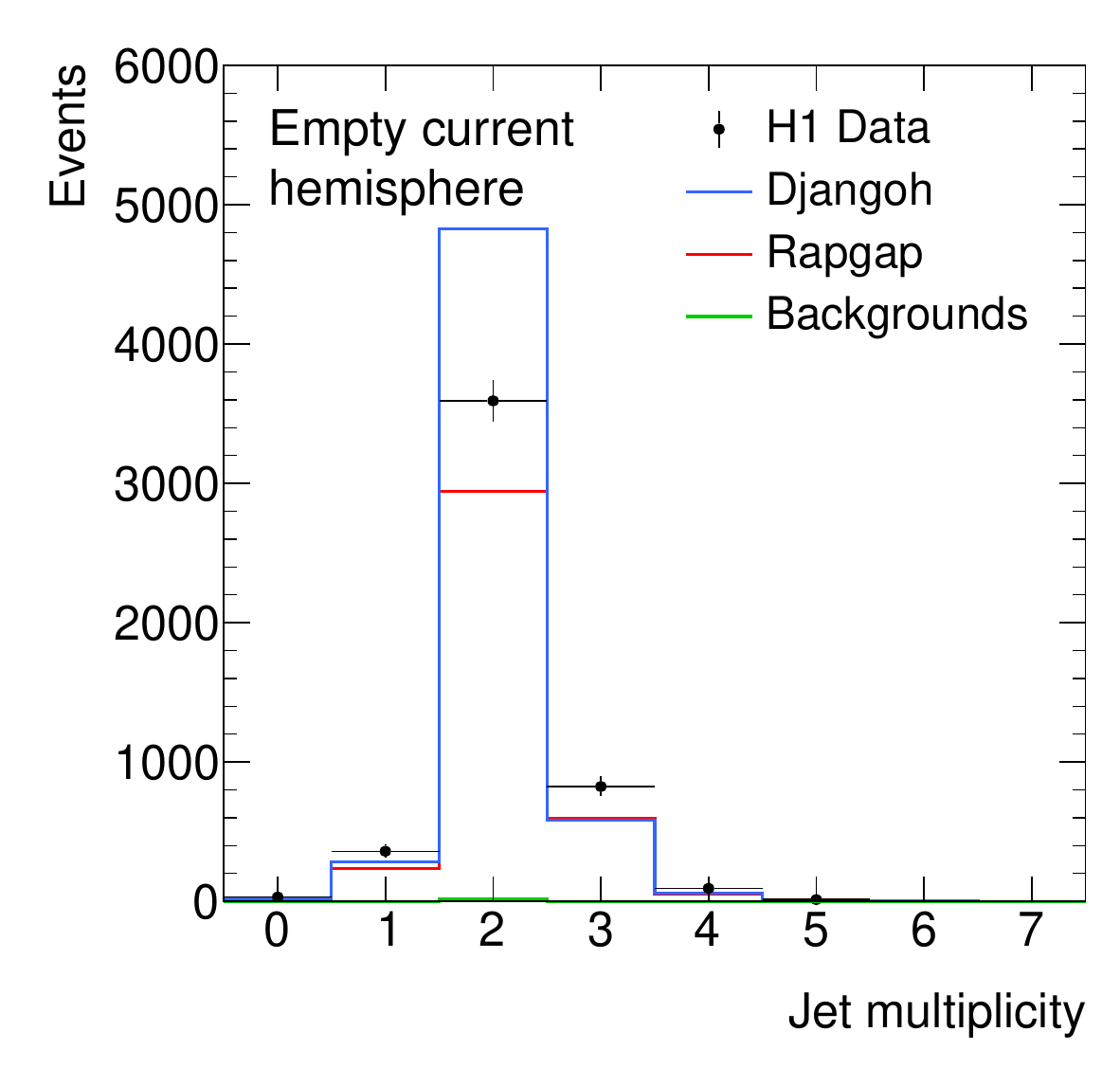
}
\caption{
    The number of jets in events with an empty current hemisphere.
    The jets are defined in the Breit frame and must exceed a transverse momentum greater than
    7\,\GeV in the Breit frame.
    Further details are given in the caption of  Fig.~\ref{fig:controlplots}.
  }
  \label{fig:njet}
\end{figure}

It is instructive to study the number of jets (the \emph{jet multiplicity}) in events with an empty hemisphere, since these events are expected to be predominantly present at $\mathcal{O}(\alpha_s)$. To this end, jets are defined in the Breit frame from all HFS objects using the $k_t$ jet algorithm~\cite{Catani:1993hr} with a distance measure of $R=1.0$ and using the $P_t$-recombination scheme.
Jets are required to have a transverse momentum $\pt$ greater than $7\,\GeV$ in the Breit frame and to fall within the acceptance of the LAr calorimeter, i.e. the polar angular range $9.4^\circ<\theta_\text{lab}^\text{jet}<154^\circ$, when boosted to the laboratory rest frame. The contributions from proton remnants or initial state radiation are highly suppressed due to limited detector acceptance. 
The jet multiplicity at the detector level is displayed in Fig.~\ref{fig:njet}.
It is observed that the jet multiplicity distribution sharply peaks at two jets and no significant number of events without jets are observed.
A few events have only one jet, which is attributed to the requirement of $P_\text{T}>7\,\GeV$, as a second jet can also be with a lower transverse momentum.

The data are corrected for acceptance and resolution effects using a
regularized matrix inversion algorithm as implemented in the TUnfold
package~\cite{Blobel:1984ku,Schmitt:2012kp}.
For each of the three observables, \xbj, $y$, and \Qsq, a detector response matrix is constructed
from the average of the Djangoh and Rapgap simulations.
The matrix contains information on migration probabilities between detector- and particle-level quantities as well as acceptance and reconstruction efficiencies.
In order to determine the fraction of empty hemisphere events, two
distributions have to be measured: the unfolded event count of empty
hemisphere events and the number of events of the inclusive DIS sample.
The response matrix therefore has a $2\times 2$ structure, differentiating events with or without an empty current hemisphere on detector or particle level. 
The two measured distributions are measured and unfolded
simultaneously and the number of DIS events is obtained at the
particle level by adding up the two respective entries.
This procedure accounts for the migrations between events with an empty current hemisphere and the ones with activities in both hemispheres.      
Additional bins at the detector level account for migrations from 
outside the DIS phase space in $\Qsq$ and $y$.
A larger number of bins than the nominal measurement bins are unfolded and the final cross sections are obtained by combining them back into nominal bins. 
These extra bins effectively divide events into smaller groups where radiation is close to the boundary of the two hemispheres. Such topologies are very sensitive to details of the QCD models and thus have to be resolved in order to minimize model uncertainties.
A very small fraction of events identified as background is subtracted from data prior to unfolding.
The contributions from photoproduction processes were estimated using Pythia~6 and the ones from the charged-current and low-\Qsq neutral current DIS processes using Djangoh~1.4.
The regularization parameter was set to $\tau=10^{-2.8}$.
It has a negligible impact on the unfolded results.
The fraction of empty hemisphere events in the inclusive DIS cross
section, $r$, is then defined as
\begin{equation}
  r :=
  \frac{n_{(E_\mathcal{C}=0)}}
       {n_{(E_\mathcal{C}>0)} + n_{(E_\mathcal{C}=0)} }
       \cdot
       c_\text{QED}
       =
       \frac{\sigma_{(E_\mathcal{C}=0)}}
            {\sigma_\text{(NC DIS)}}
       \,,
\end{equation}
where $n_i$ are the counts of events with an empty hemisphere after unfolding
($E_\mathcal{C}=0$) and the ones with activity in the current
hemisphere ($E_\mathcal{C}>0$).
The fraction $r$ then represents the ratio of the production cross
sections of empty hemisphere events to the inclusive neutral-current
DIS cross section $\sigma_\text{(NC DIS)}$. 
The factor $c_\text{QED}$ accounts for higher order QED
virtual and real corrections at the lepton vertex and QED Compton processes~\cite{H1:1999qgy,H1:2018mkk}, and
also corrects for initial lepton charge effects to an $e^-p$ initial state.
The factor  $c_\text{QED}$ is determined with Heracles as implemented in
Djangoh 1.4.
It is very close to unity, since the effects of QED
corrections as well as electroweak effects cancel in the ratio to a
large extent.

To estimate systematic uncertainties, each source of uncertainties is varied independently in simulation and the difference between the nominal results and the ones obtained from the new unfolding matrix is taken as uncertainties attributed to the source, propagated linearly to the results \cite{Schmitt:2012kp}.
Two sources of uncertainties on the energy scale of HFS objects, associated with the components contained within high \pt{} jets and the ones that are not, are treated separately by independently varying the energy of corresponding HFS objects by 
$\pm1\,\%$~\cite{H1:2014cbm}.
The angular resolutions of the azimuthal angle of the HFS objects and scattered leptons, $\pm20\,$mrad and $\pm1\,$mrad~\cite{H1:2012qti}, respectively, are taken as sources of systematic uncertainties.
The energy resolution of scattered leptons which varies from $\pm0.5\,\%$ in the backward and central regions to $\pm1\,\%$ in the forward regions is also taken into account. 
The uncertainty attributed to the
modeling of the HFS in the event generator used for
unfolding is evaluated by comparing the
results when either Djangoh or Rapgap is used to determine
the migration matrix.
This \emph{model} uncertainty accounts for acceptance, efficiency, and migration
effects due to differences in the modelling of the HFS in the MC event
generators used.
As the final cross sections are reported normalized to the inclusive
NC DIS cross section, normalization uncertainties such as luminosity
scale or trigger efficiency cancel, and several other
uncertainties including those related to the reconstruction of the scattered lepton cancel to a large extent.
For the determination of the uncertainties on $r$ after unfolding, systematic
uncertainties are considered to be fully correlated across different
bins, and statistical uncertainties have a non-zero correlation
coefficient after unfolding.

The resulting cross section ratios are compared with predictions from
Pythia~8.307~\cite{Sjostrand:2014zea,Bierlich:2022pfr,Pythia83}, which implements DIS matrix elements at leading order and various parton-shower models for higher-order emissions. Two different models are
studied: 
The default `simple' dipole-like $p_\perp$-ordered shower,
and the Dire parton
shower~\cite{Hoche:2015sya,Hoche:2017iem,Hoche:2017hno}, which is a variant of
a dipole shower model. 
Both models use the Pythia~8.3 default Lund string model for hadronization~\cite{Pythia83}.
Predictions from Powheg Box in NLO pQCD are matched to
parton shower and hadronization from Pythia
8.308~\cite{Banfi:2023mhz}.
These predictions are referred to as Powheg+Pythia.
Furthermore, predictions from the
Sherpa~2.2~\cite{Sherpa:2019gpd,Gleisberg:2008ta} and
Sherpa~3.0~\cite{Sherpa3} MC event generators are studied.
The Sherpa~2.2 predictions employ multi-leg tree-level matrix
elements~\cite{Duhr:2006iq} with up to three jets combined
with dipole showers~\cite{Catani:2001cc,Schumann:2007mg} and use a cluster hadronization model~\cite{Winter:2003tt}.
The Sherpa~3.0 predictions employ NLO QCD matrix elements from
Openloops~\cite{Buccioni:2019sur}, combined with a dipole shower based on the
truncated shower
method~\cite{Schumann:2007mg,Hoeche:2009rj,Hoeche:2012yf}.
Two hadronization models are implemented and studied: an improved cluster
hadronization model (Cluster)~\cite{Chahal:2022rid} and the string
fragmentation model of Pythia~8.3 (Lund)~\cite{Bierlich:2022pfr}.

\section{Results}
The fraction of NC DIS events with no particle candidates in the current hemisphere in
the Breit frame is determined at an $ep$ center-of-mass energy of
$\sqrt{s}=319\,\GeV$ in the kinematic region of
$150<\Qsq<1500\,\GeVsq$ and $0.14<y<0.7$.
After pre-selection, acceptance and analysis cuts, a total of 4715 event
candidates with an empty hemisphere at the detector level were recorded.
After unfolding to particle level, the fraction of NC DIS events with an
empty hemisphere in the Breit frame is measured to be
\begin{equation}
  r = 0.0112 \pm 3.9\,\%_\text{stat} \pm 4.5\,\%_\text{syst} \pm 1.6\,\%_\text{mod}\,.
  \label{eq:result}
\end{equation}
The first uncertainty represents the statistical,
the second the total experimental systematic, and the third the model uncertainty.
The fraction $r$ with its uncertaintes
is compared to various predictions in
Table~\ref{tab:tabletot1}. Only statistical uncertainties of these MC models are considered.
The predictions of Rapgap~3.1, Powheg+Pythia, Sherpa~2.2 and Sherpa~3.0 are below the data, whereas those of Djangoh~1.4 and Pythia~8.3 are above.
%
\begin{table}[bhtp!]
  \footnotesize
  \begin{tabular}{lcc}
    \toprule
          & $r$ & $\delta r$\\
    \midrule
    Data                &  $0.0112$ & $\pm 3.9\,\%_\text{stat}$ \\
                    &   & $\pm 4.5\,\%_\text{syst}$ \\
                    &   & $\pm 1.6\,\%_\text{mod}$ \\
    \midrule
    Djangoh 1.4         &  $0.0150$ & $ \pm 0.1\,\%_\text{stat}$   \\
    Rapgap 3.1          &  $0.0096$ & $ \pm 0.1\,\%_\text{stat}$ \\
    Pythia 8.3          &  $0.0127$ & $ \pm 0.1\,\%_\text{stat}$  \\
    Pythia 8.3 {\scriptsize (Dire)}  & $0.0120$ & $ \pm 0.1\,\%_\text{stat}$  \\
    Powheg+Pythia                    & $0.0107$ & $\pm 0.1\,\%_\text{stat}$  \\ 
    Sherpa 3.0 {\scriptsize(Cluster)}& $0.0100$ & $ \pm 0.1\,\%_\text{stat}$  \\
    Sherpa 3.0 {\scriptsize(Lund)}   & $0.0101$ & $ \pm 0.3\,\%_\text{stat}$  \\
    Sherpa 2.2          &  $0.00818$ & $ \pm 0.5\,\%_\text{stat}$  \\
    \bottomrule
  \end{tabular}
  \caption{  \label{tab:tabletot1}
    Comparison of the fraction $r$ of empty current hemisphere events in NC
    DIS with various predictions in the analyzed
    phase space $150<\Qsq<1500\,\GeVsq$ and $0.14<y<0.7$.
  }
\end{table}

\begin{figure*}[tbh!]
  \includegraphics[width=0.9\linewidth]{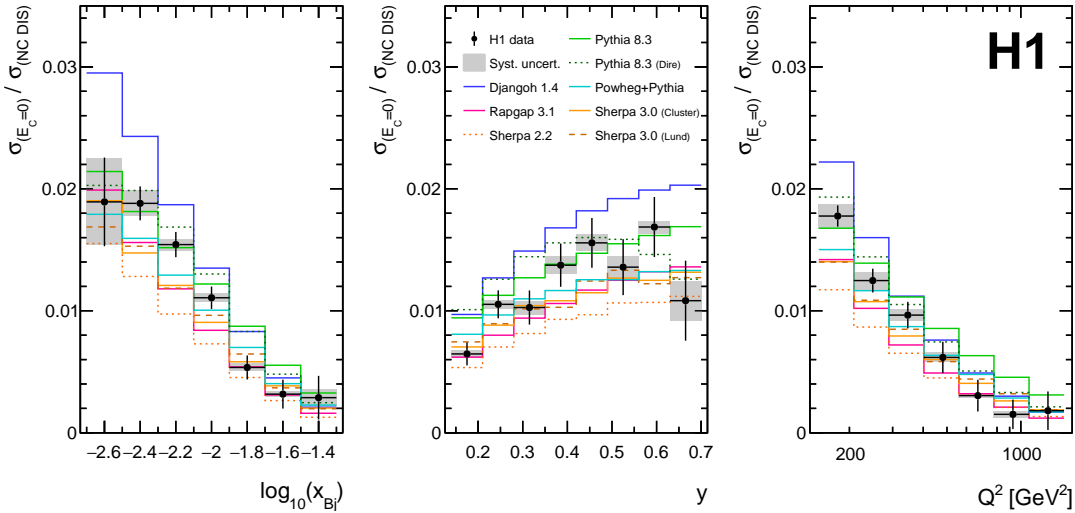}  
  \caption{
    Ratios of differential cross sections of events with an empty current
    hemisphere in the Breit frame as a function of \xbj, $y$ and \Qsq.
    The vertical bars represent the statistical uncertainties, and the shaded area the total systematic uncertainties including model uncertainties.
    Predictions from Djangoh~1.4, Rapgap~3.1, Pythia~8.3,
    Powheg+Pythia, Sherpa~2.2
    and Sherpa~3.0, are displayed with different colors and line
    styles.
    Two variants of parton shower models are studied with Pythia~8.3, and
    two hadronization models are studied with Sherpa~3.0.
  }
  \label{fig:tau1b1}
\end{figure*}

The unfolded differential fractions of empty current hemisphere events in NC DIS at the level of stable particles
are presented as function of $\log_{10}(\xbj)$, $y$ and
\Qsq\ in Table~\ref{tab:table} and are displayed in  Fig.~\ref{fig:tau1b1}.
\begin{table}[bhtp!]
  \footnotesize
  \begin{tabular}{c c @{\hskip 2em}c@{\hskip 2em} c c c}
  \toprule
  \multicolumn{2}{c}{$\log_{10}(\xbj)$} & $r$ & \multicolumn{3}{c}{Uncertainties [\%]} \\ 
  min & max &  &
    $\delta_\text{stat}$ & $\delta_\text{syst}$ & $\delta_\text{mod}$ \\ 
  \midrule
    -2.7  &     -2.5 &    0.0189 &      19 &      16 &  $   9.5$   \\
    -2.5  &     -2.3 &    0.0188 &     7.4 &     5.2 &  $   1.4$   \\
    -2.3  &     -2.1 &    0.0154 &     6.7 &     3.2 &  $  0.16$   \\
    -2.1  &     -1.9 &    0.0111 &     8.3 &     2.4 &  $  -1.8$   \\
    -1.9  &     -1.7 &   0.00535 &      18 &     5.1 &  $  -2.1$   \\
    -1.7  &     -1.5 &   0.00317 &      37 &     4.0 &  $  -5.6$   \\
    -1.5  &     -1.3 &   0.00288 &      62 &     2.0 &  $   -23$   \\
  \midrule
   \multicolumn{2}{c}{$y$} & $r$ & \multicolumn{3}{c}{Uncertainties [\%]} \\ 
   \midrule
    0.14  &     0.21 &   0.00646 &      15 &     4.1 &  $  0.1$   \\
    0.21  &     0.28 &    0.0105 &      11 &     3.0 &  $ -0.5$   \\
    0.28  &     0.35 &    0.0103 &      14 &     4.0 &  $   3.8$   \\
    0.35  &     0.42 &    0.0137 &      13 &     4.1 &  $   3.4$   \\
    0.42  &     0.49 &    0.0156 &      13 &     3.5 &  $   2.5$   \\
    0.49  &     0.56 &    0.0136 &      17 &     6.4 &  $   2.2$   \\
    0.56  &     0.63 &    0.0169 &      15 &     3.1 &  $  0.7$   \\
    0.63  &      0.7 &    0.0108 &      30 &      15 &  $  -1.5$   \\
    \midrule
   \multicolumn{2}{c}{\Qsq [\GeVsq]} & $r$ & \multicolumn{3}{c}{Uncertainties [\%]} \\ 
   \midrule
    150  &    208.4 &    0.0178 &     4.9 &     4.3 &  $   3.7$   \\
   208.4  &    289.6 &    0.0125 &     7.8 &     4.6 &  $  -3.1$   \\
   289.6  &    402.4 &   0.00965 &      11 &     4.5 &  $  -2.1$   \\
   402.4  &    559.1 &   0.00618 &      21 &     4.7 &  $  -4.0$   \\
   559.1  &    776.9 &   0.00305 &      43 &     6.5 &  $  -0.5$   \\
   776.9  &     1080 &   0.00151 &      80 &      11 &  $   -12$   \\
    1080  &     1500 &   0.00181 &      87 &     3.8 &  $  -2.2$   \\%
    \bottomrule
  \end{tabular}
  \caption{  \label{tab:table}
    Measured ratio $r$ of empty current hemisphere cross sections to the inclusive NC DIS cross sections as a function of $\log_{10}(\xbj)$, $y$, and \Qsq.
    The uncertainties specify the statistical uncertainty, the total experimental systematic uncertainty, and the model uncertainty.
  }
\end{table}
It is observed that the fraction of empty current hemisphere events decreases with increasing \xbj and \Qsq\ and with decreasing $y$. This is in line with the expectation that the phase space for Born-level two-parton topology enabling the events of interest decreases with \Qsq and \xbj.
The data are compared with predictions from Djangoh 1.4, Pythia 8.3, Powheg+Pythia,
Rapgap 3.1, Sherpa~2.2, and Sherpa~3.0.
The predictions from Djangoh and Rapgap reasonably describe the overall shape of the data, while Djangoh (Rapgap) tend to over-predict (under-predict) the data in the overall normalisation (c.f.~Table~\ref{tab:tabletot1}). The default predictions from Pythia 8.3 with default parton shower or with Dire parton shower both give a good description of the data. 
At $y \approx 0.5$, the two Pythia variants cross over and the difference is maximized at highest $y$. Unfortunately the limited data precision in this region does not clearly favor one model over the other.
The predictions from Powheg+Pythia and Sherpa~3.0 
are consistent with the data within uncertainties. 
The difference between using the cluster or the Lund string fragmentation model as implemented in Sherpa~3.0 is relatively small.
Sherpa 2.2 under-predicts the data in the entire measured \xbj and $y$ range. 

\section{Conclusions}
A measurement of the fraction of events in
neutral-current DIS, where the current hemisphere in the
Breit frame is completely empty, is presented.
The measurement is performed in the kinematic region of
$150<\Qsq<1500\,\GeVsq$ and $0.14<y<0.7$ for an electron--proton
center-of-mass energy of $319\,\GeV$.
Although the analysis of electron--proton collision events in the
Breit frame has similarities with hadronic final states in $e^+e^-$ annihilation, this
particular class of empty current hemisphere events is not present in $e^+e^-$
or $pp$ collisions.
The fraction of empty current hemisphere events was measured to be $0.0112\pm 6.2\%$ in the
selected DIS phase space.

These event configurations were predicted previously in perturbative
QCD as the first order corrections to the quark-parton model at
$\alpha_s$. 
However, the impact of higher-order pQCD corrections or hadronization
on these events has not been investigated in great
detail in the literature.

The measurement provides a first direct observation of these quite
spectacular event topologies, as well as a differential measurement as
functions of the DIS  kinematic variables \xbj, $y$ and \Qsq.
%
Among the dedicated DIS Monte Carlo event
generators Rapgap slightly underestimates the data, while
Djangoh tends to overestimate.
The general purpose event generator program Pythia 8.3 provides a
good description of the data, both with the default parton shower or the
Dire parton shower.
The event generators Powheg+Pythia and Sherpa~3.0 provide a good description of the
data, whereas Sherpa~2.2 underestimates the data.

These new data are anticipated to help improve and validate parton shower and
hadronization models for 
$e^+e^-$, $ep$ and $pp$ collision systems, and will pave the way for
optimal event generator models for future lepton--proton colliders~\cite{FCC:2018byv,LHeC:2020van,AbdulKhalek:2021gbh}.
It will be interesting to perform measurements at lower and higher \Qsq\ in the future.


\section*{Acknowledgements}
  We are grateful to the HERA machine group whose outstanding efforts
have made this experiment possible. We thank the engineers and
technicians for their work in constructing and maintaining the H1
detector, our funding agencies for financial support, the DESY
technical staff for continual assistance and the DESY directorate for
support and for the hospitality which they extend to the non-DESY members of the collaboration.
We express our thanks to all those involved in
securing not only the H1 data but also the software and working
environment for long term use,
allowing the unique H1 data set to continue to
be explored. The transfer from experiment specific to central resources with long term support,
including both storage and batch systems, has also been crucial to this enterprise.
We therefore also acknowledge the role played by DESY-IT and all
people involved during this transition and their
future role in the years to come.

\par $^{f1}$ supported by the U.S. DOE Office of Science
\par $^{f2}$ supported by FNRS-FWO-Vlaanderen, IISN-IIKW and IWT and by Interuniversity Attraction Poles Programme, Belgian Science Policy
\par $^{f3}$ supported by the UK Science and Technology Facilities Council, and formerly by the UK Particle Physics and Astronomy Research Council
\par $^{f4}$ supported by the Romanian National Authority for Scientific Research under the contract PN 09370101
\par $^{f5}$ supported by the Bundesministerium für Bildung und Forschung, FRG, under contract numbers 05H09GUF, 05H09VHC, 05H09VHF, 05H16PEA
\par $^{f6}$ partially supported by Polish Ministry of Science and Higher Education, grant DPN/N168/DESY/2009
\par $^{f7}$ partially supported by Ministry of Science of Montenegro, no. 05-1/3-3352
\par $^{f8}$ supported by the Ministry of Education of the Czech Republic under the project INGO-LG14033
\par $^{f9}$ supported by CONACYT, México, grant 48778-F
\par $^{f10}$ supported by the Swiss National Science Foundation

\bibliography{desy24-034}


\appendix

\section{Boost to the Breit frame}
\label{sec:boost}
In order to facilitate this
analysis, the matrix $M$ for the Lorentz boost from the laboratory
frame to the Breit frame was derived:
\begin{equation}
  p^\text{b} = M\cdot p^\text{Lab}\,.
  \label{eq:boostM}
\end{equation}
It can be conveniently expressed in terms of the
four-vectors $b$, Eq.~\eqref{eq:b}, and $q$.
We consider Eq.~\eqref{eq:boostM} as a plain matrix
multiplication without metric tensor. 
The laboratory rest frame is given by the right-handed HERA rest
frame, where the 
proton beam moves along the positive $z$ direction, so that the
$x$ axis points to the center of the HERA ring.
In the Breit frame we also use a right-handed coordinate system,
but adopt the definition from 
Ref.~\cite{Streng:1979pv}, such that the photon moves along the positive $z$ axis with
momentum $q^\text{b}=(0;0,0,Q)$, and thus the proton beam and the beam
fragments go along the negative $z$ axis.
The azimuthal rotation can be chosen freely.
When leaving the $x$--$y$ plane unaltered, one finds that the Lorentz boost
is expressed as
\begin{equation}
M=
-\frac{1}{Q}\begin{pmatrix}
   1 &  0 & \tfrac{Q}{\Sigma}q_x  & \tfrac{Q}{\Sigma}q_x  \\
   0 &  1 & \tfrac{Q}{\Sigma}q_y  & \tfrac{Q}{\Sigma}q_y  \\
  q_x & q_y & q_z  & -q_E  \\
  b_x & b_y & b_z  & -b_E  \\
\end{pmatrix} \,,
\end{equation}
using $\Sigma=q\cdot\hat{z}=q_E - q_z$.
The negative sign originates from the different choices of the
$z$ axis in the two frames.
When defining the $x$ axis along the transverse momentum of the
scattered lepton, the matrix $M$ becomes
\begin{equation}
M_{(\phi_{e^\prime}^\text{b}=0)} =
  - 
\begin{pmatrix}
  \tfrac{q_x}{q_\text{T}} & \tfrac{q_y}{q_\text{T}} & \tfrac{q_\text{T}}{\Sigma}  & -\tfrac{q_\text{T}}{\Sigma}  \\
  \tfrac{q_y}{q_\text{T}} & \tfrac{q_x}{q_\text{T}} & 0  & 0 \\
  \tfrac{q_x}{Q} & \tfrac{q_y}{Q} & \tfrac{q_z}{Q}  & -\tfrac{q_E}{Q}  \\
  \tfrac{b_x}{Q} & \tfrac{b_y}{Q} & \tfrac{b_z}{Q}  & -\tfrac{b_E}{Q}  \\
\end{pmatrix} \,.
\end{equation}
Hence, the longitudinal momentum and the energy component in
the Breit frame of a laboratory frame momentum $p$ is conveniently expressed as
\begin{equation}
  p_z^\text{b} = -\frac{q\cdot p}{Q} ~~~~\text{and}~~~~
  p_E^\text{b} = -\frac{b\cdot p}{Q}\,,
\end{equation}
and the current hemisphere is defined by $p^\text{b}_z>0$, while the
target/fragmentation hemisphere has $p^\text{b}_z<0$.
Interestingly, the four-vectors in the Breit frame
can be expressed without evaluating the Bjorken scaling variable
\xbj, c.f.~Eq.~\eqref{eq:b}.

\end{document}